\documentclass[12pt]{article}
\usepackage{graphicx}

\def\uiuc{Department of Physics\\
University of Illinois at Urbana-Champaign\\
1110 West Green Street, Urbana, IL, USA\\
}
\def\support{\footnote{Work supported by the National Science Foundation and the
National Center for Supercomputing Applications.}}

\def\Title#1{\begin{center} {\Large #1 } \end{center}}
\def\Author#1{\begin{center}{ \sc #1} \end{center}}
\def\Address#1{\begin{center}{ \it #1} \end{center}}

\newenvironment{Abstract}{\begin{quotation}  }{\end{quotation}}
\newenvironment{Presented}{\begin{quotation} \begin{center}
             PRESENTED AT\end{center}\bigskip
      \begin{center}\begin{large}}{\end{large}\end{center} \end{quotation}}
\def\Acknowledgements{\bigskip  \bigskip \begin{center} \begin{large}
             \bf ACKNOWLEDGEMENTS \end{large}\end{center}}

\def\beq{\begin{equation}}
\def\eeq#1{\label{#1}\end{equation}}
\def\eeqn{\end{equation}}

\def\beqa{\begin{eqnarray}}
\def\eeqa#1{\label{#1}\end{eqnarray}}
\def\eeqan{\end{eqnarray}}

\let\bar=\overbar

\def\Dslash{\not{\hbox{\kern-4pt $D$}}}
\def\dslash{\not{\hbox{\kern-2pt $\del$}}}

\def\msb{{\bar{\ssstyle M \kern -1pt S}}}

\begin{document}
\begin{titlepage}

\vfill \Title{Improved Measurement of the Muon Lifetime and
Determination of the Fermi Constant } \vfill \Author{P. T.
Debevec\support} \Address{\uiuc} \vfill
\begin{Abstract}
The MuLan collaboration has measured the lifetime of the
positve muon to a precision of 1.0~parts per million.  The
Fermi constant is determined to a precision of 0.6~parts per
million.
\end{Abstract}
\vfill
\begin{Presented}
Proceedings of CKM2010, the 6th International Workshop on the
CKM Unitarity Triangle, University of Warwick,UK, 6-10
September 2010
\end{Presented}
\vfill
\end{titlepage}
\def\thefootnote{\fnsymbol{footnote}}
\setcounter{footnote}{0}

\section{Introduction}

On July 5, 1999 the MuLan collaboration presented the proposal,
R-99-07.1: {\it A precision measurement of the positive muon
lifetime using a pulsed muon beam and the  MuLan detector}, to
the program advisory committee of the Paul Scherrer Institute
(PSI).  The proposal was to measure the positive muon lifetime
to a precision of one part per million (ppm), by which the
Fermi constant, $G_F$, would be determined to a precision of
0.5~ppm.  On December 6, 2010 the collaboration posted at
http://arxiv.org/abs/1010.0991 the manuscript of the paper,
 which reported on this measurement at the precision of 1.0~ppm.
The paper has since been published in Physical Review Letters
~\cite{MuLan:2011}. The collaboration will prepare and submit
for publication a final report of this effort.

\section{Motivation}

The Fermi constant is one of the fundamental constants of the
standard model of electroweak interactions.  The 1999 proposal
was motivated in part by the increased precision of other
electroweak parameters, e.g.~$M_Z$, and in part by the
theoretical work of Stuart and von Ritbergen
\cite{vanRitbergen:All}, who had undertaken the calculation of
two-loop QED radiative corrections, which would reduce the
theoretical uncertainty in the extraction of $G_F$ from the
muon lifetime from 15~ppm to approximately 0.2~ppm. In 1999,
the dominant error in $G_F$ was, in fact, from theory and not
from experiment.  The proposal was also motivated by the
opportunity of the intense muon beams of the PSI facility and
the opportunity of advanced pulse processing electronics and
data acquisition hardware.  In a counting experiment a
precision of one ppm requires $10^{12}$ events, a number beyond
the capabilities of previous muon lifetime measurements.

\section{Previous work}

At the same PSI advisory committee meeting in July, 1999, the
FAST collaboration presented the proposal, R-99-06.1: {\it
Precision measurement of the  $\mu^+$ lifetime ($G_F$) with the
FAST detector}, with a similar goal, but with very different
methodology. MuLan in 2007~\cite{MuLan:2007} and FAST in
2008~\cite{FAST:2008} reported 11~ppm and 16~ppm results,
respectively, from partial implementations of their respective
experiments.  Until these measurements, the 18~ppm world
average was the result from the experiment of Duclos {\it et
al.} in 1973 (140~ppm), the experiment of Balandin {\it et al.}
in 1974 (36~ppm), the experiment of Bardin {\it et al.} in 1984
(33~ppm), and the experiment of Giovanetti {\it et al.}
(27~ppm) also in 1984. These experiments separate into the two
categories of the ``radioactive-source'' mode (Duclos {\it et
al.} and Bardin {\it et al.}) implemented with a pulsed beam,
and the``one-at-a-time'' mode (Balandin {\it et al.} and
Giovanetti {\it et al.}) implemented with a low intensity
continuous beam. In the ``radioactive-source'' mode a sample of
muons is collected in a stopping medium, and their decay times
are measured from an arbitrary start time.  In the
``one-at-a-time'' mode the sample is a single muon.  The
limiting data rate of the ``one-at-a-time'' mode is a few tens
of kHz, given the 2.2~$\mu$s muon lifetime.  The limiting data
rate of the ``radioactive-source" mode is, in practice,
governed by pileup and detector stability considerations. Both
the historical and recent muon lifetime measurements are shown
in the figure.

\begin{figure}[htb]
\centering
\includegraphics[height=3.0in]{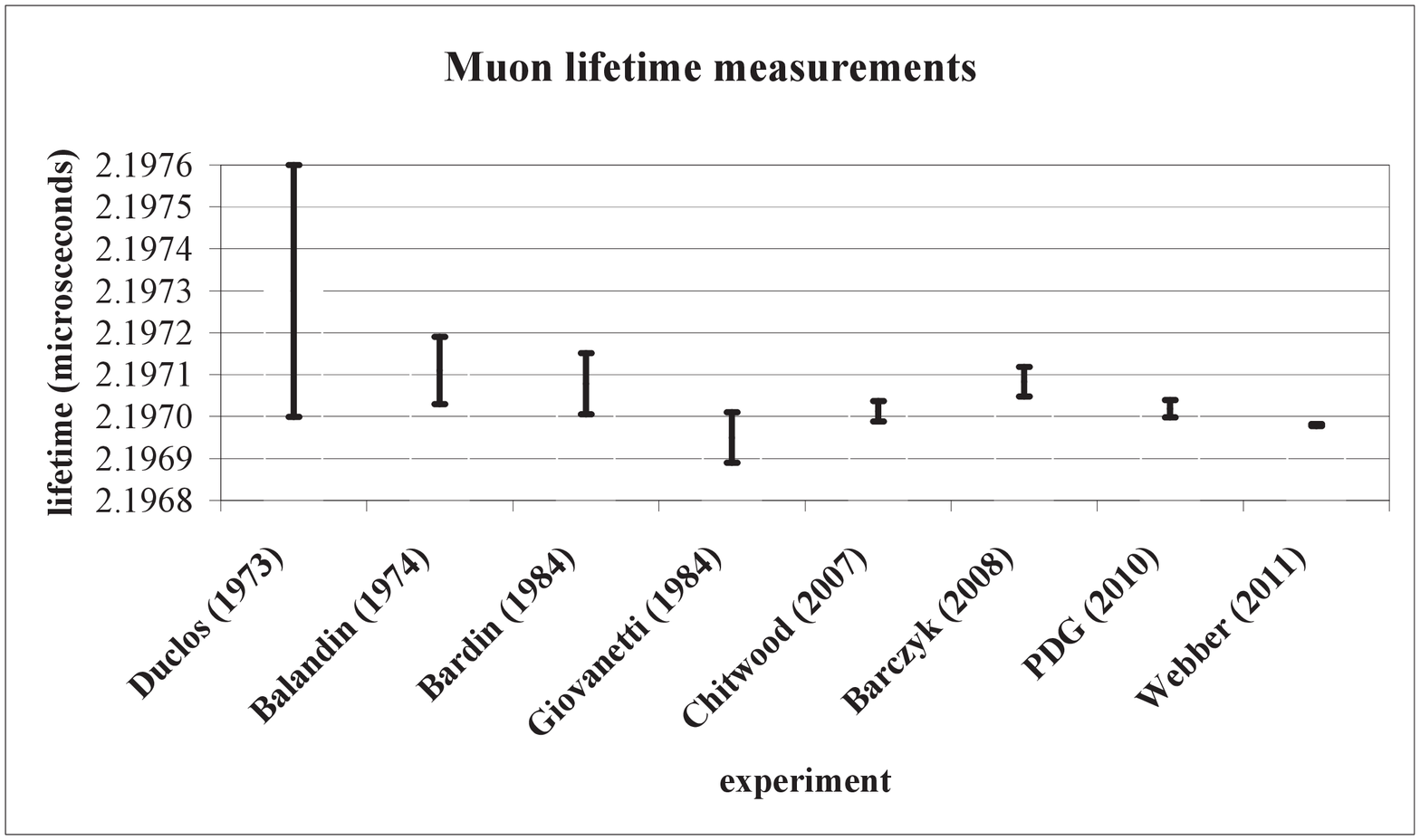}
\caption{Recent and past muon lifetime measurements.}
\label{fig:MuLan-history.eps}
\end{figure}

Statistical considerations aside, the precision of counting
experiments is controlled by systematic concerns of counting
efficiency stability (affected by both timing stability and
gain stability), background level and stability, and pileup. In
addition, the muon experiments must mitigate any effect from
muon polarization.  Muons from pion decay have unit helicity,
and muon spin relaxation and/or precession would introduce a
time dependence in the decay.  In this regard all of the
earlier experiments sought to control muon precession with
magnetic shielding of the stopping target to a level of tens of
$mG$.  In addition, all of the earlier experiments used
detectors symmetrically arranged around the stopping medium and
detectors with an effective solid angle with a large fraction
of 4~$\pi$.

\section{The MuLan experiment}

Assuming that a running time of approximately two months is
acceptable, the required data rate for a one ppm experiment is
approximately 200~kHz. Thus the ``radioactive-source'' mode
must be used at a high intensity facility. PSI has several
surface muon beams, {\it i.e.} muons produced from pions
thermalized within the production target, with appropriate
intensities.  A surface muon beam has a mean momentum of
29~MeV/c and is nearly 100~\% polarized. Because of the low
momentum, the muons must be transported in vacuum to the
stopping target.  Given the nominal 50~MHz time structure of
the PSI cyclotron, these beams are essentially CW.  The MuLan
experiment was mounted at the end of the  ${\pi}$E3 beam line,
a beam line with no permanent end station.  There is no pion
contamination in this beam line at a mean momentum of 29~MeV/c
due to its length of approximately 30~m.  There is considerable
positron contamination, but positrons are effectively removed
with a velocity-selecting $\vec{E}\times\vec{B}$ filter. The
special feature of the beam line as implemented in the MuLan
experiment is a custom, 60-ns switching, 25 kV kicker, designed
and constructed for the MuLan experiment by TRIUMF. When
energized, the muon flux of approximately 10 MHz at the target
is reduced by an extinction factor of approximately $10^{3}$.
The kicker is switched by external circuitry that synchronizes
the data collection cycles into 5~${\mu}$s kicker-off
accumulation periods, followed by 22~${\mu}$s kicker-on
measurement periods. The kicker voltage during the measurement
period determines the time stability of background from this
source.  Approximately 50~muons are collected in the
accumulation period, and approximately 20~muons remain
undecayed at the beginning of the measurement period.

Two high statistics data sets were obtained with two different
stopping targets, both chosen to mitigate the effects of muon
precession and relaxation. The stopping target in the 2006
running period was a 0.5-mm-thick foil of a ferromagnetic
alloy, Arnokrome$^{\rm TM}$ III (AK-3).  (By 2006 the kicker
was specified, designed and built, the detector and electronics
were built, the targets were chosen and fabricated, and the
beam line was understood.)  The effective internal field in
this material is approximately 0.4~T, which induces a
precession with a period of 18~ns. Thus during the accumulation
time of 5~${\mu}$s the polarization of the muon ensemble is
reduced by a factor of approximately $10^3$, as the muons
arrive randomly and precess. The stopping target in the 2007
running period was a 2-mm-thick disk of crystalline quartz, in
which stopped muons form muonium 90~\% of the time.  A Halbach
arrangement of permanent magnets provides a nearly uniform
$130~G$ field in the plane of the quartz disk. In this field
muonium precesses with a period of 2.6~ns.  The 10~\% muon
component in diamagnetic states precesses with a period of
550~ns, which is observed, since the average number of muon
rotations during the accumulation period is not large.  (Note,
however, that the sum of events recorded by a detector at angle
$(\theta, \phi)$ with those from an equally efficient detector
at $(180^\circ - \theta, 180^{\circ}+\phi)$ form a decay
histogram that is immune to precession and relaxation.  A
detector with point symmetry, viewing a point source, is immune
to the effect of muon spin precession and relaxation.)

The detector array consists of 170 stacked pairs of 3-mm-thick
plastic scintillators. They are arranged in a truncated
icosahedron geometry and grouped in 20 hexagon and 10 pentagon
assemblies.  With this segmentation the probability that a pair
records an electron during the measurement period is 10~\%.
Allowing for the beam entry and exit ports and gaps between
detector pairs, the detector encompasses 75~\% of 4$\pi$  and
is point symmetric about the center of the stopping target. The
center of a pair of detectors is 38.3~cm from the target. Each
scintillator is viewed by either a Photonis or Electron Tubes
29-mm photomultiplier.  On average, 80 photoelectrons are
registered for each minimum-ionizing particle.  The signals
from the 340 photomultiplier tubes are recorded using 450~MHz,
8-bit waveform digitizers.  The waveform digitizer sampling
frequency is set by an Agilent E440 signal generator having
stability better than 0.01~ppm/month.  The frequency was set to
$\pm 450$~ppm of 451~MHz, with the exact value unknown to the
collaboration and only disclosed after the completion of a
blind analysis of the two datasets.  Thus the photomultiplier
waveforms were sampled at a time interval of 2.2~ns (one
``clock tick'').  Normally 24~ADC samples (53~ns) comprise a
full waveform; they are recorded when an input signal exceeds a
set threshold with four to eight samples preceding the trigger
point.  The digitization is extended if the threshold is
exceeded at the 24th sample.  For the average pulse, the full
width at 20~\% maximum is 9~ns.

\section{Data analysis}

The 130 TB of raw waveforms are converted into lists of valid
``hits''. The waveforms are fit to pulse-shape templates,
prepared from a set of low-rate events. The time of a decay is
defined as the peak of the pulse shape.  In more than 99.9~\%
of the cases, a single pulse exists on a waveform, and it can
be identified reliably.  Rare multiple-pulse waveforms are fit
by an iterative approach which adds additional pulses as
needed. The fitting procedure works reliably when two pulses
are separated by more than three ``clock ticks.''  For shorter
separation times, only one hit is reconstructed and its time is
set to the pulse-height weighted average of the ADC samples.
For resolved pulses, an ``artificial deadtime'' (ADT) can be
applied on a per-detector basis eliminating pulses when a
minimum time separates sequential hits in the same
scintillator.  The ADT used in the analysis varies from 5 to 68
``clock ticks'' and represents an important diagnostic of the
pileup correction procedure.

Histograms are filled with coincident events from each detector
pair; the coincidence window interval is set to the ADT.  For
the AK-3 data set, the muon lifetime is obtained from a fit to
the sum of the 170 individual histograms using the three
parameter function
\begin{equation}
F(t) = A\exp(-t/\tau_\mu) + B, \label{eq:fit}
\end{equation}
where $B$ accounts for the flat background.  A fit to the
pileup-corrected event histogram, summed over all detector
pairs, gives a   $\chi^2/{\rm dof} = 1.03 \pm 0.04$.  For the
quartz data set, in which slow precession and relaxation of the
diamagnetic muon component is present, each detector pair
histogram is first fit with $A$ multiplied by the function
$[1+P_2\cdot\exp{(-t/\tau_2)}\sin(\omega t+\phi)]$; where $P_2$
and $\tau_2$  are transverse polarization and relaxation
parameters.  The precession and relaxation parameters vary
smoothly over the highly uniform detector, and the muon
lifetime is obtained from a fit to this distribution. The
leading systematic uncertainty in this fitting procedure is
from the uncertainty of the beam position on the target. The
lifetime obtained in this manner is in agreement to 0.3~ppm
with the result from a fit to the simple sum of all
detector-pair histograms without regard to precession and
relaxation. Despite the complication of muon spin precession
and relaxation, a reliable muon lifetime is obtained.

Pileup is corrected using a statistical procedure based on the
data itself.  Basically, each measurement period presents an
identical random sequence of decays.  When a hit is observed at
a time $t_i$ in fill $j$, an interval between $t_i$ and $t_i
+$~ADT is searched in fill $j+1$.  If a hit is observed in this
interval, its time is recorded in a separate histogram, which
is then added back into the original decay histogram.  This
process is repeated for higher-order pileup and a Monte-Carlo
study with the full statistics of the experiment verified the
pileup correction procedure.  Pileup is found to contribute a
systematic error of 0.20~ppm.  Gain and timing stability were
examined at a similar sensitivity and were found to contribute
to the systematic error at a level of 0.25~ppm and 0.12~ppm.
The total systematic uncertainty for each running period was
0.42~ppm.

The stability of the lifetime versus the starting time of the
fit is a powerful diagnostic because pileup, gain and time
stability, and spin precession and relaxation effects might all
exhibit time dependencies.  For both the 2006 and 2007 data
sets the extracted lifetime does not depend on the fit start
time, apart from the statistically allowed variation.

\section{Results}

The results for the two running periods are in excellent
agreement:

\begin{eqnarray}\nonumber
\tau_{\mu}({\rm R06})&=& 2196979.9 \pm 2.5 \pm 0.9 {\rm ~ps}, \\
\tau_{\mu}({\rm R07})&=& 2196981.2 \pm 3.7 \pm 0.9 {\rm ~ps}.
\end{eqnarray}

As is customary, the first error is statistical and the second
is systematic.  Combining the two results we obtain

\begin{equation}
\tau_{\mu}({\rm MuLan}) = 2196980.3 \pm 2.2 {\rm ~ps~~~(1.0~ppm)},
\label{finalresult}
\end{equation}

The MuLan result is more than 15 times as precise as any other
individual measurement and consequently dominates the world
average.  The MuLan result lies $2.5~\sigma$  below the current
PDG average.

Following the theoretical framework of Stuart and van Ritbergen
and the significant emendation of Pak and Czarnecki\cite{Pak},
we obtain the most precise determination of the Fermi constant:

\begin{equation}
G_F({\rm MuLan}) = 1.1663788(7) \times 10^{-5} {\rm
~GeV^{-2}}~~~({\rm 0.6~ppm}).
\end{equation}

The Particle Data Group has not (yet) adopted the
abovementioned methodology so $G_F$(MuLan) should not be
directed compared to $G_F$(PDG2010), which was obtained with a
different methodology.

\Acknowledgements No undertaking of this scope can succeed
without significant assistance from many parties.  The MuLan
collaboration acknowledges the efforts of our PSI colleagues,
especially D.~Renker, K.~Deiters, M.~Hildebrandt, and
U.~Rohrer, and our TRIUMF colleagues, M.~Barnes and G.~Wait.
PTD acknowledges a crucial discussion with W. Marciano at the
CKM2010 workshop.  The National Center for Supercomputer
Applications at the University of Illinois hosted the data
analysis effort, and the National Science Foundation provided
financial support to all of the institutions of the MuLan
collaboration.

\end{document}